\documentclass[12pt]{article}
\usepackage{amsmath,amssymb,bm,graphicx}
\usepackage{cite}
\usepackage{color} 
\usepackage{hyperref}

\newcommand{\pslash}{\not \! p}

\newcommand{\delslash}{\not \! \partial}

\begin{document}

\vskip 0.5 truecm

\begin{center}
{\Large{\bf Naturalness in 
see-saw mechanism\\\vskip 0.3cm and Bogoliubov transformation}}
\end{center}
\vskip .5 truecm
\begin{center}
{\bf { Kazuo Fujikawa$^{1,2}$ and Anca Tureanu$^1$}}
\end{center}

\begin{center}
\vspace*{0.4cm} 
{\it {$^1$Department of Physics, University of Helsinki, P.O.Box 64, 
\\FIN-00014 Helsinki,
Finland\\
$^2$Quantum Hadron Physics Laboratory, RIKEN Nishina Center,\\
Wako 351-0198, Japan
}}
\end{center}

%\makeatletter
%\@addtoreset{equation}{section}
%\def\theequation{\thesection.\arabic{equation}}
%\makeatother

\begin{abstract}
We present an alternative perspective on the see-saw mechanism for the neutrino mass, according to which the small 
neutrino mass is given as a difference of two large  masses. 
This view emerges when  an analogue of the Bogoliubov transformation is used to describe Majorana neutrinos in the Lagrangian of the see-saw mechanism, which is analogous to the BCS theory. 
The Bogoliubov transformation clarifies the natural appearance of Majorana fermions  when C is strongly violated by the right-handed neutrino mass term with good CP in the single flavor model.  
Analyzing typical models with $m_{R}$= $10^{4}$ to $10^{15}$ GeV, it is shown that a hitherto unrecognized fine tuning of the order $m_{\nu}/m_{R}=10^{-15}$ to $10^{-26}$ is  required to make the commonly perceived see-saw mechanism work in a natural setting, namely, when none of the dimensionless coupling constants are very small. 
 
\end{abstract}
%\maketitle
%\large
 
\section{Introduction}

When one discusses the natural appearance of the observed very small neutrino masses~\cite{particledata}, one often refers to the see-saw mechanism~\cite{minkowski,yanagida, mohapatra} the precise form of which depends on specific models~\cite{review}. Those models are characterized by a very large mass scale and thus the natural appearance of the tiny neutrino mass is rather surprising. Naturalness is an esthetic notion and thus subjective, and it should ultimately be determined by experiments.  Currently active search for the support of the see-saw mechanism in the form of Majorana neutrinos is going, and we expect that this esthetical issue will be tested soon by experiments.

  It may also be appropriate to examine the naturalness of the see-saw mechanism from a different perspective. 
We attempt to understand the natural appearance of the eigenstates of charge conjugation C, Majorana fermions, using an analogue of the Bogoliubov transformation when C is strongly violated by the right-handed neutrino mass term which has good CP symmetry.  We then recognize that the tiny neutrino mass in the see-saw mechanism is given as a difference of two large masses, precise values of which depend on models.
This suggests a view different from the conventional one,  motivating us to ask whether the see-saw mechanism  is "natural" in the sense emphasized, for example, in~\cite{weinberg, susskind}.  We show that a hitherto unrecognized  fine tuning of the order $m_{\nu}/m_{R}$ is required to make the see-saw mechanism work in a natural setting.  
 
We first recapitulate the basic properties of Majorana fermions, namely, charge conjugation and parity.
The Majorana fermions are defined by the condition 
$$\psi(x)=C\bar{\psi}^{T}(x)=\psi^{c}(x),$$
where $C=i\gamma^{2}\gamma^{0}$ stands for the charge conjugation 
matrix~\cite{bjorken}; the quantity $C\bar{\psi}^{T}(x)$ is directly evaluated for given $\psi(x)$ while $\psi^{c}(x)$ is evaluated by a unitary charge conjugation operator, and the agreement of these two expressions provides an important consistency check in our analysis of, for example,  eq. \eqref{majorana} below. 
We start with a generic neutral Dirac fermion, which is denoted by $\nu(x)$ for later convenience, and define the combinations
$$\psi_{\pm}(x)=\frac{1}{\sqrt{2}}[\nu(x)\pm \nu^{c}(x)],$$
which satisfy
$$\psi^{c}_{\pm}(x)=\pm\psi_{\pm}(x),$$
showing that $\psi_{+}(x)$ and $\psi_{-}(x)$ are Majorana fields. We treat the 
fermion with $\psi^{c}_{-}(x)=-\psi_{-}(x)$ also as a Majorana fermion.

It is well-known~\cite{bjorken,weinberg2} that,  in theories where the fermion number is conserved, 
 discrete symmetries such as parity can generally be defined  with an arbitrary phase freedom $\delta$,
$$\nu(x)\rightarrow e^{i\delta}\gamma^{0}\nu(t,-\vec{x}).$$
The conventional parity $\nu(x)\rightarrow \gamma^{0}\nu(t,-\vec{x})$ and $\nu^{c}(x)\rightarrow -\gamma^{0}\nu^{c}(t,-\vec{x})$   for the Dirac fermion (in the following called "$\gamma^{0}$-parity") corresponds to $\delta=0$ and thus satisfying $P^{2}=1$.
One can confirm that parity for an {\em isolated single} Majorana fermion is consistently defined only by "$i\gamma^{0}$-parity" with $\delta=\pi/2$, i.e. $\nu(x)\rightarrow i\gamma^{0}\nu(t,-\vec{x})$ and $\nu^{c}(x)\rightarrow i\gamma^{0}\nu^{c}(t,-\vec{x})$,
namely by (see Ref.~\cite{weinberg2})
\begin{eqnarray}\label{2}
\psi_{\pm}(x)\rightarrow i\gamma^{0}\psi_{\pm}(t,-\vec{x}).
\end{eqnarray}
This definition is consistent with the reality of $\psi_{\pm}(x)$ in the Majorana representation, where $\gamma^{0}$ is hermitian but purely imaginary. 
The phase freedom $\delta$ is thus fixed by the Majorana condition and ${\rm P}^{2}=-1$. We are interested in Majorana fermions, therefore we exclusively use this "$i\gamma^{0}$-parity" in this paper.

\section{Model Lagrangian and Bogoliubov transformation}

We analyze the hermitian Lorentz invariant quadratic Lagrangian for a single flavor of the neutrino, which is a minimal extension of the Standard Model,
\begin{eqnarray}\label{3}
{\cal L}&=&\overline{\nu}_{L}(x)i\gamma^{\mu}\partial_{\mu}\nu_{L}(x)+\overline{n}_{R}(x)i\gamma^{\mu}\partial_{\mu}n_{R}(x)\nonumber\\
&-&m\overline{\nu}_{L}(x)n_{R}(x)
-(m_{L}/2)\nu_{L}^{T}(x)C\nu_{L}(x)\nonumber\\
&-&(m_{R}/2)n_{R}^{T}(x)Cn_{R}(x) + h.c.,
\end{eqnarray}
where $n_{R}(x)$ is a right-handed analogue of $\nu_{L}(x)$, and $m$, $m_{L}$, and $m_{R}$ are real parameters. We define a new Dirac-type variable
\begin{eqnarray}\label{4}
\nu(x)\equiv \nu_{L}(x) + n_{R}(x)
\end{eqnarray}
in terms of which the above Lagrangian is re-written as
\begin{eqnarray}\label{5}
{\cal L}&=&(1/2)\{\overline{\nu}(x)[i\delslash - m]\nu(x)+ \overline{\nu^{c}}(x)[i\delslash - m]\nu^{c}(x)\}\nonumber\\
&-&(\epsilon_{1}/4)[\overline{\nu^{c}}(x)\nu(x) +\overline{\nu}(x)\nu^{c}(x)]\nonumber\\
&-&(\epsilon_{5}/4)[\overline{\nu^{c}}(x)\gamma_{5}\nu(x) -\overline{\nu}(x)\gamma_{5}\nu^{c}(x)],
\end{eqnarray}
where $\epsilon_{1}=m_{R}+m_{L}$ and $\epsilon_{5}=m_{R}-m_{L}$.
The C and P transformation rules for $\nu(x)$ are defined by 
\begin{eqnarray}\label{7}
\nu^{c}(x)=C\bar{\nu}^{T}(x), \ \ \nu^{p}(x)=i\gamma^{0}\nu(t,-\vec{x}), 
\end{eqnarray}
and thus $\nu(x) \leftrightarrow \nu^{c}(x)$ under C and $\nu^{c}(x)\rightarrow i\gamma^{0}\nu^{c}(t,-\vec{x})$ under P; CP is given by 
\begin{eqnarray} \label{CP}
\nu^{cp}(x)=i\gamma^{0}C\bar{\nu}^{T}(t,-\vec{x}).
\end{eqnarray}
The above  Lagrangian \eqref{5} is CP conserving, although C and P ($i\gamma^{0}$-parity) are separately broken by the last term. 

In defining Majorana fermions, the exact meaning of the charge conjugation operation C is crucial. In literature (see, e.g., Ref.~\cite{review}), one customarily defines the charge conjugation in the Lagrangian \eqref{3} by 
\begin{eqnarray}\label{8}
(\nu_{L}(x))^{c}=C\bar{\nu}_{L}^{T}(x), \ \ (n_{R}(x))^{c}=C\bar{n}_{R}^{T}(x).
\end{eqnarray}
We must emphasize that the symbols $(\nu_{L}(x))^{c}$ and $(n_{R}(x))^{c}$ are not to be understood as "transformation laws" but rather as mnemonics for the quantities on the right-hand side, since a unitary operator to generate those transformations does  not exist. This can be clearly seen by the following contradictions. If one assumes the action of the unitary charge conjugation operator, one has $\nu_{L}(x)=[(1-\gamma_{5})/2]\nu_{L}(x)$ and 
%\begin{eqnarray}\label{c-conjugate}
$$(\nu_{L}(x))^{c}={\cal C}\nu_{L}(x){\cal C}^{\dagger}=[(1-\gamma_{5})/2]{\cal C}\nu_{L}(x){\cal C}^{\dagger}=[(1-\gamma_{5})/2]C\bar{\nu}_{L}^{T}(x),$$
%\end{eqnarray}
 which imply $(\nu_{L}(x))^{c}=0$, and similarly for $n_{R}(x)$. 
Moreover, the well-known C- and P-violating weak interaction Lagrangian is written as 
\begin{eqnarray}\label{weak-int}
{\cal L}_{W}&=&(g/\sqrt{2})\bar{e}_{L}\gamma^{\mu}W^{(-)}_{\mu}(x))\nu_{L}+ h.c.\nonumber\\
&=&(g/\sqrt{2})\bar{e}_{L}\gamma^{\mu}W^{(-)}_{\mu}(x))[(1-\gamma_{5})/2]\nu_{L}+ h.c. .
\end{eqnarray}
If one assumes again \eqref{8} as transformation laws, the first expression implies that ${\cal L}_{W}$ is invariant under C, while the second expression implies ${\cal L}_{W}\rightarrow 0$. CP (or CPT)  is the only reliable way to define a chiral antiparticle. More comments on this issue will be given later. 

The transformation rules  \eqref{7} for the Lagrangian \eqref{5} are operatorially well defined, and they imply 
\begin{eqnarray}\label{9}
\nu^{c}_{L,R}(x)=\left(\frac{1\mp \gamma_{5}}{2}\right)\nu^{c}(x)=C\overline{\nu}^{T}_{R,L}(x),
\end{eqnarray}
as well as 
\begin{eqnarray}\label{parity-2}
\nu^{p}_{L,R}(x)=i\gamma^{0}\nu_{R,L}(t,-\vec{x}),
\end{eqnarray}
 namely, {\em doublet representations} of C and P  for $\nu_{L}(x)$ and $n_{R}(x)$, which are not symmetries of \eqref{3} for $m_{L}\neq m_{R}$. The CP transformation 
\begin{eqnarray}\label{CP-2}
\nu^{cp}_{L,R}(x)=i\gamma^{0}C\overline{\nu}^{T}_{L,R}(t,-\vec{x})
\end{eqnarray}
 is an exact symmetry of the original Lagrangian \eqref{3}.  We thus adopt the Lagrangian \eqref{5} and the (unitary) C and P transformations \eqref{7} as the basis of our analysis, which 
defines a prototype of the Lagrangian of the see-saw mechanism~\cite{minkowski, yanagida, mohapatra, review} for $m_{L}\simeq 0$, where the right-handed Majorana-type mass $m_{R}$ is added to the Dirac fermion with mass $m$. An analogy of the Lagrangian  \eqref{5} with the Bardeen--Cooper--Schrieffer (BCS) theory was noted some time ago~\cite{chang}.

To solve \eqref{5}, we apply an analogue of Bogoliubov transformation, $(\nu, \nu^{c})\rightarrow (N, N^{c})$, defined as
\begin{eqnarray}\label{10}
\left(\begin{array}{c}
            N(x)\\
            N^{c}(x)
            \end{array}\right)
&=& \left(\begin{array}{c}
            \cos\theta\, \nu(x)-\gamma_{5}\sin\theta\, \nu^{c}(x)\\
            \cos\theta\, \nu^{c}(x)+\gamma_{5}\sin\theta\, \nu(x)
            \end{array}\right),
\end{eqnarray}
with
$\sin 2\theta =(\epsilon_{5}/2)/\sqrt{m^{2}+(\epsilon_{5}/2)^{2}}$.
We can then show that the anticommutators are preserved, i.e.,
\begin{eqnarray}\label{anti-comm}
&& \{N(t,\vec{x}), N^{c}(t,\vec{y})\}=\{\nu(t,\vec{x}), \nu^{c}(t,\vec{y})\},\nonumber\\  
&&\{N_{\alpha}(t,\vec{x}), N_{\beta}(t,\vec{y})\}=\{N^{c}_{\alpha}(t,\vec{x}), N^{c}_{\beta}(t,\vec{y})\}=0,
\end{eqnarray}  
and thus it satisfies the canonicity condition of the Bogoliubov transformation. A transformation analogous to \eqref{10} has been  successfully used in the analysis of neutron-antineutron oscillations \cite{FT}.

After the Bogoliubov transformation, which diagonalizes the Lagrangian with $\epsilon_{1}=0$,  ${\cal L}$ in \eqref{5} becomes
\begin{eqnarray}\label{13}
{\cal L}&=&\frac{1}{2}\left[\overline{N}(x)\left(i\delslash - M\right)
 N(x)+\overline{N^{c}}(x)\left(i\delslash - M\right)N^{c}(x)\right]\nonumber\\
&-&\frac{\epsilon_{1}}{4}[\overline{N^{c}}(x)N(x) + \overline{N}(x)N^{c}(x)],
\end{eqnarray}
with the mass parameter
\begin{eqnarray}\label{14}
M\equiv \sqrt{m^{2}+(\epsilon_{5}/2)^{2}}.
\end{eqnarray}
This implies that the Bogoliubov transformation maps the original theory to
a theory characterized by the new large mass scale $M$ ($\epsilon_{5}/2$ corresponds to the energy gap). The Bogoliubov  transformation maps a linear combination of a Dirac fermion and its charge conjugate to another Dirac fermion, and thus the Fock vacuum is mapped to a new orthogonal vacuum defined by ${\cal L}$ in \eqref{13} with $\epsilon_{1}=0$ at $t=0$~\cite{chang}.  It is important that the Bogoliubov transformation \eqref{10} preserves the CP symmetry,
although it does not preserve the transformation properties under $i\gamma^{0}$-parity and C separately.
In the present single flavor model, this leads to the Lagrangian \eqref{13} of the Bogoliubov quasi-fermion $N(x)$ which is symmetric under the $i\gamma^{0}$-parity and C transformations. 

The Lagrangian \eqref{13} is exactly diagonalized by 
\begin{eqnarray}\label{diagonal} 
\psi_{+}(x)=\frac{1}{\sqrt{2}}(N(x)+N^{c}(x)),\ \ \ 
\psi_{-}(x)=\frac{1}{\sqrt{2}}(N(x)-N^{c}(x)),
\end{eqnarray}           
in the form
\begin{eqnarray}\label{16}
{\cal L}&=&\frac{1}{2}\{\overline{\psi}_{+}[i\delslash-M_{+}]\psi_{+}%\nonumber\\
%&&\ 
+\overline{\psi}_{-}[i\delslash-M_{-}]\psi_{-}\},
\end{eqnarray}
with the masses 
\begin{eqnarray}\label{masses}
M_{\pm}=M \pm \epsilon_{1}/2,
\end{eqnarray}
and $m_{\nu}=M_{-}$ corresponds to the small neutrino mass.
The charge conjugation and $i\gamma^{0}$-parity properties,  which are required for the isolated massive Majorana fermions, 
\begin{eqnarray}\label{17}
&&\psi^{c}_{\pm}(x)=\pm \psi_{\pm}(x), \ \  \psi^{p}_{\pm}(x)= i\gamma^{0}\psi_{\pm}(t,-\vec{x}),
\end{eqnarray}
are thus consistent with  the transformation properties of $N(x)$.

In the terminology of the familiar "mixing matrix", \eqref{10} is regarded as a transformation between the "mass eigenstate" $(N, N^{c})$ (the transformation from $N$ to $\psi_{\pm}$ is symmetry-wise trivial) and the "flavor eigenstate" $(\nu, \nu^{c})$; the mass eigenstate in Minkowski space is characterized by the full Lorentz symmetry including P and T, and thus C because of CPT, while the flavor eigenstate is constrained by the original Lagrangian \eqref{5}, which is not invariant under those operations.
The original neutrino is expressed in terms of the Majorana fermions $\psi_{\pm}$ if one uses \eqref{10} as 
\begin{eqnarray}\label{19}
\nu(x)&=&[(\cos\theta +\sin\theta \gamma_{5})/\sqrt{2}]\psi_{+}(x)+[(\cos\theta -\sin\theta \gamma_{5})/\sqrt{2}]\psi_{-}(x),
\nonumber\\
\nu^{c}(x)&=&[(\cos\theta -\sin\theta \gamma_{5})/\sqrt{2}]\psi_{+}(x)-
[(\cos\theta +\sin\theta \gamma_{5})/\sqrt{2}]\psi_{-}(x),
\end{eqnarray}
  but 
the unitary C operations on $\psi_{\pm} \rightarrow \pm \psi_{\pm}$ in the expression of $\nu(x)$ do not reproduce $\nu^{c}(x)$ in \eqref{19}, reflecting the C breaking in the original Lagrangian \eqref{5}.  The Bogoliubov transformation thus  explains the natural appearance of Majorana fermions in the C-breaking Lagrangian \eqref{5}.   

We here comment on the effects of CP breaking on see-saw mechanism. The most general Lagrangian is defined by real $m$ and $\epsilon_{5}$ and  complex
$\epsilon_{1}e^{i\alpha}$ (correspondingly $\epsilon_{1}e^{-i\alpha}$ in the second term with $\epsilon_{1}$) in \eqref{5} after a suitable choice of the phase of $\nu$. Then  $\alpha\neq 0$ implies CP breaking. The exact mass spectrum is given by $M_{\pm}=\left([M\pm \sqrt{\epsilon^{2}_{1}-\tilde{\epsilon}^{2}_{1}}/2]^{2}+(\tilde{\epsilon}_{1}/2)^{2}\right)^{1/2}$ with $\tilde{\epsilon}_{1}=\epsilon_{1}\sin\alpha\sin 2\theta$ if one uses the Bogoliubov transformation (see Ref.~\cite{FT}), and thus the parameter $\alpha$ which modifies the neutrino mass is an observable. CP violation in \eqref{3} and \eqref{5} is restricted to be very small, $\alpha\sim m_{L}/m_{R}< 10^{-15}$, to have a successful see-saw mechanism in \eqref{30} below, in the present single flavor model. 
  
\section{See-saw mechanism}

It is customary to discuss the see-saw mechanism in perturbation theory.
One may re-write the Lagrangian \eqref{5} with $m=0$ and $\epsilon_{1}=\epsilon_{5}=m_{R}$, which is diagonalized  as 
\begin{eqnarray}\label{20}
{\cal L}&=&(1/2)[\overline{\tilde{\psi}}_{+}(x)i\gamma^{\mu}\partial_{\mu}\tilde{\psi}_{+}(x) - m_{R}\overline{\tilde{\psi}}_{+}(x)\tilde{\psi}_{+}(x)]\nonumber\\
&+&(1/2)[\overline{\tilde{\psi}}_{-}(x)i\gamma^{\mu}\partial_{\mu}\tilde{\psi}_{-}(x)],
\end{eqnarray}
in terms of the fields defined by
\begin{eqnarray}\label{21}
            \left(\begin{array}{c}
            \nu(x)\\
            \nu^{c}(x)
            \end{array}\right)
            &=&\left(\begin{array}{c}
            \frac{1+\gamma_{5}}{2}\tilde{\psi}_{+}(x)+\frac{1-\gamma_{5}}{2}      
            \tilde{\psi}_{-}(x)\\
            \frac{1-\gamma_{5}}{2}\tilde{\psi}_{+}(x)-\frac{1+\gamma_{5}}{2}      
            \tilde{\psi}_{-}(x)            
            \end{array}\right).
\end{eqnarray} 
The CP symmetry of $\tilde{\psi}_{\pm}$,
\begin{eqnarray}\label{22}
\tilde{\psi}^{cp}_{\pm}(x)=(\pm i\gamma^{0})\tilde{\psi}_{\pm}(t,-\vec{x}),
\end{eqnarray}
is consistently translated to the CP symmetry of $\nu$ and $\nu^{c}$ in \eqref{21}. Note that the  Lagrangian \eqref{5} with $m=0$ and $\epsilon_{1}=\epsilon_{5}=m_{R}$ is CP invariant but not C invariant.
One may then perform a second-order perturbation analysis by treating
the Dirac mass term $m$ in \eqref{5} as a small perturbation. To be explicit,
\begin{eqnarray}\label{mass-term}
{\cal L}_{I}=-\frac{m}{2}[\overline{\nu}\nu+\overline{\nu^{c}}\nu^{c}]=\frac{m}{2}[\overline{\tilde{\psi}}_{+}\gamma_{5}\tilde{\psi}_{-}-\overline{\tilde{\psi}}_{-}\gamma_{5} \tilde{\psi}_{+}].
\end{eqnarray}
One then obtains the second order perturbative result, symbolically,
\begin{eqnarray}\label{25}
 (m^{2}/2!)\overline{\tilde{\psi}}_{-}\gamma_{5}\langle T\tilde{\psi}_{+}\overline{\tilde{\psi}}_{+}\rangle\gamma_{5} \tilde{\psi}_{-}
\simeq (-i/2)(m^{2}/m_{R})\overline{\tilde{\psi}}_{-}
\tilde{\psi}_{-},
\end{eqnarray}
using $\langle T\tilde{\psi}_{+}\overline{\tilde{\psi}}_{+}\rangle=\frac{i}{\pslash-m_{R}}$ near on-shell $\pslash=0$ of $\tilde{\psi}_{-}$. This mass term is added to the massless fermion $\tilde{\psi}_{-}$ in \eqref{20}. The massless fermion thus acquires a mass $m^{2}/m_{R}$ and still satisfies the CP-conjugation property \eqref{22}, being still a Majorana fermion. This mechanism to generate a small neutrino mass $m^{2}/m_{R}$ is called {\em see-saw mechanism}, and the essence of the see-saw mechanism is to
deal with {\em a ratio of two numbers}~\cite{minkowski,yanagida,mohapatra, review}.

One can confirm that  $i\gamma^{0}$-parity transformation $\tilde{\psi}^{p}_{\pm}(x)= i\gamma^{0}\tilde{\psi}_{\pm}(t,-\vec{x})$ and charge conjugation
$\tilde{\psi}^{c}_{\pm}(x)= \pm \tilde{\psi}_{\pm}(x)$ of Majorana fermions {\em do not} induce the corresponding parity and charge conjugation of $\nu$ and $\nu^{c}$ in \eqref{21}, just as in \eqref{19}.
Also, one can confirm from \eqref{21} that 
\begin{eqnarray}\label{majorana}
\tilde{\psi}_{+}(x)=\nu_{R}+ C\bar{\nu_{R}}^{T},\ \ \ \tilde{\psi}_{-}(x)=\nu_{L}-C\bar{\nu_{L}}^{T},
\end{eqnarray} 
and thus the above {\em failure} of C is related to the inconsistency of  $\nu^{c}_{L,R}=C\bar{\nu}^{T}_{L,R}$ in \eqref{8} when regarded as transformation laws. Our message is that we identify $\tilde{\psi}_{\pm}(x)$ as Majorana fermions  {\em not} because of the unjustified relation $\nu^{c}_{L,R}=C\bar{\nu}^{T}_{L,R}$ but because of CP symmetry in \eqref{21} and \eqref{22}, and the relation  \eqref{21} shows the C violation in the original Lagrangian \eqref{5}. In contrast,  we have shown that the variables $\psi_{\pm}(x)=(N\pm N^{c})/\sqrt{2}$  
in  \eqref{16} satisfy the well-defined charge conjugation property \eqref{17}
using the charge conjugation property of 
the massive Bogoliubov quasi-fermion $N$ in the new vacuum.

\section{Naturalness}

We have obtained the exact solution $\nu(x)$ of \eqref{5} in terms of the well-defined massive Majorana fermions $\psi_{\pm}(x)$ as in \eqref{17}.
By tentatively setting $\epsilon_{1}=\epsilon_{5}=m_{R}$ (namely,
$m_{L}=0$),  we have
\begin{eqnarray}\label{mixing-angle}
\sin2\theta= (m_{R}/2)/\sqrt{m^{2}+(m_{R}/2)^{2}},
\end{eqnarray}
and the Bogoliubov quasi-fermion $N(x)$ is defined in the mass scale $M=\sqrt{m^{2}+(m_{R}/2)^{2}}$. The mass spectrum of $\psi_{\pm}(x)$ is given by $M_{\pm}=M\pm m_{R}/2$  in \eqref{masses},
in particular,
\begin{eqnarray}\label{28}
M_{-}=M- m_{R}/2
=\frac{m^{2}}{\sqrt{m^{2}+(m_{R}/2)^{2}}+ m_{R}/2},
\end{eqnarray}
namely, the tiny neutrino mass $m_{\nu}=m^{2}/[\sqrt{m^{2}+(m_{R}/2)^{2}}+ m_{R}/2]$ on the right-hand side is given by a {\em difference} of two gigantic masses.
We thus encounter potential naturalness issues related with the neutrino mass, and we discuss below the possible implications of \eqref{28}.

The quadratic divergence in  the Higgs masses, which is related to the see-saw mechanism, has been analyzed in the past (see, e.g., Ref. \cite{review}).  The fermion masses have no quadratic divergences and thus appear to have no direct difficulty of that type. We however emphasize that  the Higgs masses and the fermion masses are both related to the vacuum expectation values of Higgs fields. If the potential of Higgs fields should be modified by radiative corrections, the fermion masses rewritten in terms of Higgs vacuum values defined as a stationary point of the renormalized effective potential are inevitably modified as is indicated by the Coleman--Weinberg mechanism~\cite{coleman}. The ordinary argument of the multiplicative renormalization of fermion masses and their stability under renormalization is valid in spontaneously broken gauge theory only when the renormalization of the Higgs potential is well-controlled.

Supersymmetry has been expected to resolve issues related to the quadratic divergence in  the Higgs mass and also the general issues of hierarchy.  In view of no obvious indication yet of supersymmetry at LHC, one may think of possible alternatives to supersymmetry.  The remarkable success of the Standard Model may suggest that some forms 
of the scaling argument proposed, for example, in Ref.~\cite{bardeen} are working. As a resolution of the issue of quadraric divergences in such a scheme, one may argue for the generic nature of the dimensional regularization (i.e., a suitably formulated higher derivative regularization  reproduces the main results of the dimensional regularization)~\cite{fujikawa}, which is also related to the derivation of the Callan--Symanzik equation~\cite{callan, symanzik} without encountering quadratic divergences~\cite{callan}.  The fact that the quadratic divergence  is not important in a properly formulated
 Wilsonian renormalization group flow has also been shown in Ref.~\cite{iso}, for example.  Our view is that the quadratic divergence itself is not a major issue but the issue of hierarchy remains.
 
The esthetic aspect of the naturalness issue, which is our main interest,  is related to a difference of two
large numbers to define a small number. The common argument about the Higgs mass, when a naive cut-off is applied~\cite{susskind}, is that we have symbolically $({\rm Planck\ mass})^{2} - ({\rm Planck\ mass})^{2} = ({\rm Higgs\ mass})^{2}$,
or $10^{38} \mbox{GeV}^{2} -10^{38} \mbox{GeV}^{2} = 10^{4} \mbox{GeV}^{2}$. Here the fine tuning ratio is
$\sim 10^{4}/10^{38}=10^{-34}$, although the subtraction of a large number by itself is consistent with the basic idea of renormalization. In our analysis, we assume the dimensional regularization and thus the issue of the quadratic divergence
in the Higgs mass itself does not arise.

In the case of the see-saw mechanism, \eqref{28} shows symbolically that ``GUT'' mass $-$ ``GUT'' mass $=$ neutrino mass.  Here ``GUT'' mass stands for a generic mass much larger than the Standard Model mass scale; a natural choice (see Ref.~\cite{review}) appears to be for
$m_{R}$ in the range $10^{4}\ \mbox{GeV}$ to $10^{15}\ \mbox{GeV}$ with $m$ in the range of electron mass to top quark mass, to generate the observed values of neutrino masses by $\sim m^{2}/m_{R}$.
We thus have $10^{4}\ \mbox{GeV}-10^{4}\ \mbox{GeV}= 10^{-2}\  \mbox{eV}$ if one adopts $m_{R}=10\ \mbox{TeV}$, and the fine tuning ratio  is 
$\sim 10^{-2}/10^{13}=10^{-15}$; if one adopts $m_{R}=10^{15}\ \mbox{GeV}$ this ratio becomes $\sim 10^{-26}$.  In the present paper, we assume that all the fermion masses in the starting Lagrangian are generated by the Higgs mechanism. We can then write the ratio in the form  (adopting the specific value $m_{R}=10\ \mbox{TeV}$ in the following),
\begin{eqnarray}\label{extra}
m_{\nu}/m_{R}=m^{2}/m^{2}_{R}\simeq v^{2}/D^{2}<10^{-15}
\end{eqnarray}
by assuming the natural (approximately) universal Yukawa couplings for the vacuum value $v$ of the ordinary Higgs  and  the vacuum value $D$ of an extra scalar, which generates $m_{R}$.  We assume the gauge invariant dimensional regularization, as already stated, which eliminates  quadratic divergences and clearly separates the issue of quadratic divergence from the issue of hierarchy~\footnote{We thank Luis \'Alvarez-Gaum\'e for a clarifying discussion on the importance of separating the issue of hierarchy from the issue of quadratic divergence.  The quadratic divergence is removed by the dimensional regularization but the issue of hierarchy remains.}. The  relation \eqref{extra} then leads to a hitherto unrecognized  interesting fine tuning as sketched below.
 
\section{Fine tuning in the see-saw mechanism}
 
 In the analysis of the see-saw mechanism defined by \eqref{5}, the pure right-handed case, $\epsilon_{1}-\epsilon_{5}=2m_{L}=0$, is often assumed~\cite{bilenky}.  
 The simplest possibility to realize such a case is to add a gauge singlet massive $n_{R}$ to the Standard Model, together with a Dirac mass term. The Majorana-type mass for $n_{R}$ is added by hand and thus our argument of fine tuning is not relevant in this case~\footnote{If $m$ is of the order of top quark mass and thus $m_{R}$ is very large, we have a fine tuning of the order $m^{2}_{R}-m^{2}_{R}=v^{2}$ by analyzing the self-energy of the Higgs boson arising from a fermion loop.}.
 Such a choice may however  be tantamount to an arbitrary adjustment of the neutrino mass to observed values. Besides, the Dirac-type neutrino  in such a scheme without adding the right-handed mass term may enjoy more enhanced lepton number symmetry in the sense of the naturalness argument of {}'t Hooft~\cite{thooft}. 

We thus discuss the schemes which generate the right-handed mass term in a non-trivial way.
 For definiteness, we 
consider a concrete model with gauge group $SU(2)_{L}\times SU(2)_{R}\times U(1)$  in the original proposal~\cite{minkowski} and the case with the  left-handed as well as right-handed neutrino masses as in \eqref{5}.  The model contains three Higgs multiplets with quantum numbers of the above gauge group specified by (see Ref.~\cite{minkowski})
\begin{eqnarray}\label{higgs}
\varphi_{L}: (3, 1, \mp 2), \ \ \varphi_{R}: (1,3; \pm 2),\
 \  \varphi_{LR}: (2, 2; 0),
 \end{eqnarray}
 where the $SU(2)_{R}$ triplet $\varphi_{R}$ generates $m_{R}$ and $\varphi_{LR}$ contains the ordinary Higgs doublet, and fermion doublets,
 \begin{eqnarray}\label{fermion}
 l_{L}= \left(\begin{array}{c}
            \nu(x)\\
            e(x)
            \end{array}\right)_{L}, \ \ \  
l_{R}= \left(\begin{array}{c}
            n(x)\\
            e(x)
            \end{array}\right)_{R}.
 \end{eqnarray}
 This model is closely related to the model in~\cite{mohapatra} with emphasis on different aspects, and also the generalizations of this model encompass many of the interesting models of see-saw mechanism~\cite{yanagida, review}.
 We  also rewrite the mass formula $m_{\nu}=M_{-}$ as
\begin{eqnarray}
m_{\nu}&=&\sqrt{m^{2}+(\epsilon_{5}/2)^{2}}- \epsilon_{1}/2\nonumber\\
& \simeq& m_{R}(\epsilon_{5}-\epsilon_{1})/(\epsilon_{5}+\epsilon_{1})+m^{2}/m_{R}
\end{eqnarray}
for $\epsilon_{1}\gg m$ and $\epsilon_{5}\gg m$, but both being close to $m_{R}$ in \eqref{16}. We thus have to satisfy 
\begin{eqnarray}\label{30}
|(\epsilon_{5}-\epsilon_{1})/(\epsilon_{5}+\epsilon_{1})|
<m_{\nu}/m_{R}\leq 10^{-15},
\end{eqnarray}
to make the see-saw mechanism work (by adopting $m_{R}=10$ TeV). 
 
 In this setting, using the Bogoliubov transformation, we identify the fine tuning in the form of a very accurate parity violation 
\begin{eqnarray}\label{parity-violation} 
|(\epsilon_{5}-\epsilon_{1})/(\epsilon_{5}+\epsilon_{1})|=m_{L}/m_{R}<10^{-15}. 
 \end{eqnarray}
  In the picture of Bogoliubov quasi-fermions,
the starting theory has the very large mass scale $M$ as in \eqref{13} and the neutrino mass is given by the enormous cancellation of large masses. This suggests that the fine tuning is a relevant issue.  
Correspondingly,
 in the concrete model~\cite{minkowski}, one may choose the vacuum values of Higgs fields tuned such that $\epsilon_{1}-\epsilon_{5}\simeq 0$ (i.e., $\langle \varphi_{L}\rangle=\langle S\rangle\simeq 0$ in eq. (10) in~\cite{minkowski}) is satisfied.  
 At each order of perturbation theory, one needs to impose or fine-tune the condition  of small $m_{L}$ (namely, $\langle \varphi_{L}\rangle\simeq 0$) even if the potential and thus vacuum values of Higgs fields receive sizable corrections. This is illustrated below by first analyzing the related  Type II see-saw model.

We recall  the Type II see-saw model~\cite{review, ma}  where one analyzes the coupling of  an $SU(2)_{L}$ triplet scalar boson $\Delta$ with a large mass $M_{\Delta}$ to the left-handed neutrino through a Yukawa coupling $ Y_{\Delta}\nu_{L}^{T}C\Delta\nu_{L}$, and a triple scalar coupling $\lambda_{\Delta}M_{\Delta}H^{2}\Delta$  to the Standard Model Higgs $H$. The tadpole-type diagram (i.e., one of the ends of $\Delta$-propagator landing on the neutrino line and the Higgs vacuum value is attached to the other end of the propagator) or  a direct analysis of the tree level full potential~\cite{ma} then leads to the left-handed  neutrino mass $m_{L}=\lambda_\Delta Y_{\Delta}v^2/M_\Delta$, where $v$ is the vacuum value  of $H$.
 This neutrino mass term may be compared with Weinberg's dimension five operator~\cite{weinberg3} if one chooses $M_{\Delta}$ around the GUT mass scale such as $\sim 10^{15}$ GeV.  The fine tuning in this model appears if one considers a one-loop tadpole correction (with heavy $\Delta$-field drawing a loop) to the quartic coupling $\lambda_{3}\Delta^{2}H^{2}\rightarrow \lambda_{3}M_{\Delta}^{2}\ln(M_{\Delta}/v)H^{2}$ using the dimensional regularization~\cite{fujikawa}. To maintain the Standard Model Higgs mechanism with the ordinary vacuum value $v$ so that the neutrino mass $m_{L}$ is kept small, one needs to satisfy $|\lambda_{3}M_{\Delta}^{2}|<v^{2}$, namely, $|\lambda_{3}|<v^{2}/M_{\Delta}^{2}=10^{-26}$, which is also disposed of by  a fine tuning of the induced term {\em minus} a finite local counter term in the Higgs mass term $m^{2}_{H}H^{2}$, expressed by the symbolic notation $M_{\Delta}^{2}-M_{\Delta}^{2}=v^{2}$, without making $|\lambda_{3}|$ very small. Note that this extra fine tuning is not required if one does not attempt to generate a small $m_{L}$ by introducing a large $M_{\Delta}$.

One can write the triple scalar coupling  in the above Type II model as $\lambda_{\Delta}DH\Delta H\\ \simeq \lambda_{\Delta}(m_{R}/Y_{\Delta})H\Delta H$, which is confirmed to be natural in the context of the model in~\cite{minkowski}  if one identifies $D=\langle \varphi_{R}\rangle$ and $\Delta=\varphi_{L}$ and $H$ with the first column of $\varphi_{LR}$ in \eqref{higgs}.
We then have the left-handed neutrino mass $m_{L}\simeq \lambda_{\Delta}m_{R}v^{2}/M_{\Delta}^{2}$ and thus
$\lambda_{\Delta}v^{2}/M_{\Delta}^{2}\simeq m_{L}/m_{R}<10^{-15}$ as in \eqref{parity-violation} for  the successful (conventional) see-saw  mechanism in~\cite{minkowski}.  When $|\lambda_{\Delta}|$ is not very small, $M_{\Delta}^{2}$ becomes very large and we  need a fine tuning  in the symbolic notation $M_{\Delta}^{2}-M_{\Delta}^{2}=v^{2}$ in the Higgs mass term by analyzing a one-loop tadpole with heavy $\varphi_{L}$ drawing a loop in the quartic coupling $\lambda_{3}{\rm Tr}\varphi_{L}^{\dagger}\varphi_{L} {\rm Tr}\varphi_{LR}^{\dagger}\varphi_{LR}\rightarrow \lambda_{3}M_{\Delta}^{2}\ln(M_{\Delta}^{2}/v^{2}) {\rm Tr}\varphi_{LR}^{\dagger}\varphi_{LR}$, as in the case of the above Type II model.  This fine tuning keeps the tree level $v^{2}$
unchanged so that the tiny $m_{L}$ is kept small without being disturbed by the one-loop correction.  

The tree level condition $|\lambda_{4}|<v^{2}/D^{2}< 10^{-15}$ with $D=\langle \varphi_{R}\rangle$ needs to be satisfied as suggested by \eqref{extra} for the quartic scalar  coupling $\lambda_{4}
{\rm Tr}\varphi_{R}^{\dagger}\varphi_{R} {\rm Tr}\varphi_{LR}^{\dagger}\varphi_{LR}$ of an $SU(2)_{R}$ triplet $\varphi_{R}$, if one wants to maintain the conventional Higgs mechanism.  Instead of making $|\lambda_{4}|$ very small, this condition is replaced by a tuning  of the {\em tree-level} Higgs mass term $m_{LR}^{2}{\rm Tr}\varphi_{LR}^{\dagger}\varphi_{LR}$ to absorb $D^{2}$ in a symbolic notation $D^{2}-D^{2}=v^{2}$.   At the one-loop level,  the tadpole with heavy $\varphi_{R}$ drawing a loop induces $\lambda_{4}
{\rm Tr}\varphi_{R}^{\dagger}\varphi_{R} {\rm Tr}\varphi_{LR}^{\dagger}\varphi_{LR}\rightarrow \lambda_{4}M_{\varphi_{R}}^{2}\ln(D^{2}/v^{2}){\rm Tr}\varphi_{LR}^{\dagger}\varphi_{LR}$ using the dimensional regularization~\cite{fujikawa}.  Thus the  fine tuning in the symbolic notation  $D^{2}-D^{2}=v^{2}$ in the Higgs mass term is  required at the one-loop level also, by noting $M_{\varphi_{R}}\sim D$ in the model~\cite{minkowski}. This fine-tuning keeps $v^{2}$ unchanged after the one-loop correction and thus keeps the neutrino masses $m_{\nu}$ and $m_{L}$ small.  Note again that this extra fine tuning is not required if one does not attempt to make $m_{R}$ large by introducing large $D=\langle \varphi_{R}\rangle$  to generate a small $m_{\nu}$. 

The present analysis implies that some form of {\em extra} fine tuning (or extra sizable finite renormalization) is inevitable to explain the tiny neutrino mass using very heavy particles in a {\em natural} setting (namely, if one starts with the assumption that none of the dimensionless coupling constants are very small); the degree of fine tuning is given by $m_{\nu}/m_{R}$ in \eqref{extra}.  We mainly discussed the original model in~\cite{minkowski}, but our analysis, which is related to  an analysis of hierarchy problem where a heavy particle makes a light particle heavy through quantum corrections, is applicable to the related model in~\cite{mohapatra} and also to other interesting models of see-saw mechanism~\cite{yanagida, mohapatra, review} which are regarded as generalizations of the model we analyzed. 
The generalization of the present analysis to the case of three generations of leptons contains some technical complications, but the basic observation remains valid, namely, the Bogoliubov transformation maps the model to another model characterized by a large mass scale and the ratio $m_{\nu}/m_{R}$ provides a degree of fine tuning in the bosonic sector.

\section{Discussion and Conclusion}

The Bogoliubov transformation nicely  explains the appearance of Majorana fermions in the theory with strong C violation in the single flavor model and it suggests a hitherto unrecognized fine tuning of the order $m_{\nu}/m_{R}=m^{2}/m_{R}^{2}$. In our analysis it is crucial to recognize that the stability argument of the fermion sector and Higgs sector under quantum corrections is inseparably connected as is indicated by the Coleman-Weinberg mechanism~\cite{coleman}. Naturalness is an esthetic notion and thus subjective, but we believe that our analysis provides useful information when one appreciates the simple version of the see-saw mechanism as perceived by the majority of physicists. 
As for the practical aspects of the see-saw mechanism, one may consider many refinements such as the ``very low-energy'' see-saw scheme with $m_{R}<1 $ keV~\cite{gouvea},  for example, which may have  interesting physical implications.  Those refinements of the see-saw mechanism and their applications are beyond the scope of the present paper.   If SUSY should be discovered below the energy scale of GUT, our fine tuning argument, which is related to the hierarchy issue,  is substantially modified~\cite{luis}.   
\\

We thank Masud Chaichian for very helpful discussions on all the aspects of this paper. We also thank Robert Shrock  and Luis \'Alvarez-Gaum\'e for clarifying discussions. 
This work is supported in part by the Vilho,Yrj\"o and Kalle V\"ais\"al\"a Foundation. The support of the Academy of Finland under the
Projects no. 136539 and 272919 is gratefully acknowledged.


\begin{thebibliography}{99}
\bibitem{particledata}
K.A. Olive et al. (Particle Data Group), Chin. Phys. C {\bf 38} (2014) 090001.
\bibitem{minkowski}
P. Minkowski, Phys. Lett. B{\bf 67} (1977) 421.
\bibitem{yanagida}
T. Yanagida, in Proceedings of Workshop on Unified Theory and Baryon Number
in the Universe, ed. by O. Sawada and A. Sugamoto (KEK report 79-18,
1979), p. 95.\\
M. Gell-Mann, P. Ramond and R. Slansky, in Supergravity, ed. by P. van
Nieuwenhuizen and D.Z. Freedman (North-Holland, Amsterdam, 1979), p. 315.
\bibitem{mohapatra}
R. N. Mohapatra and G. Senjanovic, Phys. Rev. Lett. {\bf 44}   
 (1980) 912 .
\bibitem{review}
For a review of various schemes in the see-saw mechanism, see, for example, Zhi-zhong Xing,
"Neutrino Physics", in Proceedings of the 1st Asia-Europe-Pacific School of High-Energy Physics (AEPSHEP), 2012, published in CERN Yellow Report CERN-2014-001,177-217, arXiv:1406.7739 [hep-ph].
\bibitem{weinberg}
S. Weinberg,  Phys. Rev. D{\bf 19} (1979) 1277.
\bibitem{susskind}
L. Susskind, Phys. Rev. D{\bf 20} (1979) 2619.
\bibitem{bjorken}
J.D. Bjorken and S. D. Drell, {\em Relativistic Quantum Fields}  (McGraw Hill, New York, 1965).
\bibitem{weinberg2}
S. Weinberg, {\em The Quantum Theory of Fields} I 
(Cambridge University Press, Cambridge, England, 1995).
\bibitem{chang}
L.N. Chang and N.P. Chang, Phys. Rev. Lett. {\bf 45} (1980) 1540.
\bibitem{FT}
K. Fujikawa and A. Tureanu,  Phys.\ Rev.\ D {\bf 94} (2016) 115009,
arXiv:1609.03203.
\bibitem{coleman}
S. Coleman and E.J. Weinberg, Phys. Rev. D{\bf 8} (1973) 1888.
\bibitem{bardeen}
W.A.  Bardeen, "On naturalness in the standard model", FERMILAB-CONF-95-391-T.
\bibitem{fujikawa}
K. Fujikawa, Phys. Rev. D{\bf 83} (2011) 105012; "Dimensional regularization is generic", arXiv: 1605.05813.
\bibitem{callan}
C.G. Callan, Phys. Rev. D{\bf 2} (1970) 1541.   
\bibitem{symanzik}
K. Symanzik, Commun. Math. Phys. {\bf 18} (1970) 227.
\bibitem{iso}
H. Aoki and S. Iso, Phys. Rev. D{\bf 86} (2012) 013001. 
\bibitem{bilenky}
S. Bilenky, {\em Introduction to the Physics of Massive and Mixed Neutrinos},      Lect. Notes. Phys. {\bf 817} (Springer, Berlin Heidelberg, 2010).
\bibitem{thooft}
G.  {}'t Hooft, ``Naturalness, chiral symmetry, and spontaneous chiral symmetry breaking''
in {\em Recent Developments in Gauge Theories}, Cargese 1979, eds. G.  {}'t Hooft et
al. (Plenum, New York, 1990).
\bibitem{ma}
E. Ma and U. Sarkar, Phys. Rev. Lett. {\bf 80} (1998) 5716.
\bibitem{weinberg3}
S. Weinberg, Phys. Rev. Lett. {\bf 43} (1979) 1566.
\bibitem{gouvea}
A. de Gouvea, Phys.Rev. D{\bf 72} (2005) 033005.\\
A. de Gouvea, J. Jenkins, N. Vasudevan, Phys. Rev. D{\bf 75} (2007) 013003. 
\bibitem{luis}
L. \'Alvarez-Gaum\'e and M. \'A. V\'azquez-Mozo, {\em An Invitation to Quantum Field Theory}  (Springer, Berlin Heidelberg, 2012).
\end{thebibliography}
\end{document}